      [1999/12/01 v1.4c Il Nuovo Cimento]
\documentclass{cimento}

\title{Kaon-nuclei interaction studies at low energies (the AMADEUS project)}
\author{K.~Piscicchia\from{ins:x}\from{ins:evil}\thanks{E-mail: kristian.piscicchia@lnf.infn.it}\ETC,
M.~Bazzi\from{ins:x},
C.~Berucci\from{ins:x},
D.~Bosnar\from{ins:settimo},
A.M.~Bragadireanu\from{ins:secondo},
M.~Cargnelli\from{ins:terzo},
A.~Clozza\from{ins:x},
C.~Curceanu\from{ins:x},
A.~D'Uffizi\from{ins:x},
F.~Ghio\from{ins:quinto},
C.~Guaraldo\from{ins:x},
P.~Kienle\from{ins:terzo}\from{ins:sesto},
M.~Iliescu\from{ins:x},
T.~Ishiwatari\from{ins:terzo},
P.~Levi Sandri\from{ins:x},
J.~Marton\from{ins:terzo},
D.~Pietreanu\from{ins:x},\from{ins:secondo},
M.~Poli Lener\from{ins:x},
A.~Rizzo\from{ins:x},
A.~Romero Vidal\from{ins:x},
E.~Sbardella\from{ins:x},
A.~Scordo\from{ins:x},
D.L.~Sirghi\from{ins:x},\from{ins:secondo},
F.~Sirghi\from{ins:x},\from{ins:secondo},
H.~Tatsuno\from{ins:x},
I.~Tucakovic\from{ins:x},
O.~Vazquez Doce\from{ins:ottavo},
E.~Widmann\from{ins:terzo},
J.~Zmeskal\from{ins:terzo},}
\instlist{\inst{ins:x} INFN Laboratori Nazionali di Frascati, Frascati (Roma), Italy \inst{ins:evil} Dipartimento di Fisica, Universita' degli Studi 'Roma Tre', Roma, Italy  \inst{ins:settimo} Physics Department, University of Zagreb, Zagreb, Croatia \inst{ins:secondo} IFIN-HH, Magurele, Bucharest, Romania   \inst{ins:terzo} Stefan-Meyer-Institut f$\ddot{u}$r subatomare Physik, Vienna, Austria     \inst{ins:quinto} INFN Sez. di Roma I and Inst. Superiore di Sanita', Roma, Italy  \inst{ins:sesto}  Technische Universit$\ddot{a}$t M$\ddot{u}$nchen, Physik Department, Garching, Germany \inst{ins:ottavo} Excellence Cluster Universe, Technische Universit$\ddot{a}$t M$\ddot{u}$nchen
Boltzmannstr. 2, D-85748, Garching, Germany}
\PACSes{\PACSit{13.75.Jz}{Kaon-baryon interactions}
\PACSit{25.80.Nv}{Kaon-induced interactions}\PACSit{21.65.Jk}{Mesons in nuclear matter}}
\begin{document}

\maketitle

\begin{abstract}
The AMADEUS experiment aims to perform dedicated precision studies in the sector of low-energy kaon-nuclei interaction at the DA$\Phi$NE collider at LNF-INFN. In particular the experiment plans to perform measurements of the debated deeply bound kaonic nuclear states (by stopping kaons in cryogenic gaseous targets $^3$He and $^4$He) to explore the nature of the $\Lambda(1405)$ in nuclear environment and to measure the cross section of K$^-$ on light nuclei, for K$^-$ momentum lower than 100 MeV/c.
The AMADEUS dedicated setup will be installed in the
central region of the KLOE detector
%The physics program and the proposed setup (consisting of dedicated additional items inserted in the
%central region of the KLOE detector) are presented. Preliminary results from the analysis of the
%existing KLOE data and future plans will be discussed.
\end{abstract}

\section{Introduction}
The AMADEUS (Antikaon Matter At DA$\Phi$NE Experiments with Unraveling Spectroscopy) experiment \cite{ref:AMADE,ref:AMADEUS}  will study the low energy interactions of kaons with nucleons and nuclei. AMADEUS aims to search for the most fundamental Deeply Bound Kaonic Nuclear States (DBKNS), that are the kaonic dibaryon states ($K^-pp\;,\;K^-pn$), produced by stopping $K^-$ in a $^3$He target and, as a next step, the kaonic tribaryon states ($K^-ppn\;,\;K^-pnn$) using a $^4$He target. A great experimental effort is ongoing in the search for such states, but the possible experimental indications of the formation of dibaryonic and tribaryonic states, have received alternative explanations in the framework of known processes \cite{Oset:2005sn}.
Another important subject is the $\Lambda(1405)$ resonance \cite{ref:hyji}, whose nature is still not completely
understood. The structure of the $\Lambda(1405)$ has been found to be important in various aspects in the strangeness sector of nonperturbative QCD. AMADEUS will give the possibility to better understand such state and its behavior in nuclear environment, with high statistics.
Moreover, AMADEUS plans to perform the measurement of low energy cross sections of
charged kaons on H, d, and He (for kaons momentum lower than 100 MeV/c) and the study of
nuclear interactions of K$^-$ in various targets.

\section{Setup performance requirements}
Kaonic nuclear states will be studied both in the formation process, by means of missing mass spectroscopy, and in the decay process. As DBKNS are expected to decay into states containing $\Lambda$ and $\Sigma$ hyperons, neutral (neutrons and photons) as well as charged particles are to be detected, with momenta in a wide range. All these detection requirements are perfectly satisfied by
the KLOE detector, which is made of a 4$\pi$ cylindrical drift chamber (DC) and a calorimeter, both
immersed in the 0.52 T field of a superconducting solenoid. KLOE has an acceptance of 96$\%$, is
optimized for detection of charged particles in the relevant energy range, and has good detection
efficiency for neutrons, as was checked by the KloNe group \cite{ref:anelli}. On the other side, DA$\Phi$NE is a
unique source of low energy kaons. DA$\Phi$NE is a double ring e$^+$ e$^-$ collider, designed to work in the
center of mass energy of the $\phi$  meson. Charged kaons coming from $\phi$ decay are characterized by low momentum
($\sim$ 127 MeV/c), which enables to stop them in gaseous targets, and a back to back topology which
turns to be ideal for triggering purpose.

\section{The dedicated AMADEUS setup}

The AMADEUS setup will be implemented inside the KLOE DC, between the beam pipe ($6$ cm diameter) and the DC entrance wall ($50$ cm diameter). Two main components of the experimental setup are presently under development: a high density cryogenic gaseous target and a trigger system.

An essential feature of the detector is the possibility to trigger on charged kaons coming from
the interaction point. The main goal is a time resolution sufficient to clearly distinguish Kaons from background. 
This will be achieved by making use of two layers of scintillating
fibers read at both sides by silicon photomultipliers (SiPM) surrounding the beam pipe. Employing two high granularity layers will
give the possibility to perform a preliminary tracking as well.
SiPM turn to be optimal for our purposes as they are rather insensitive to magnetic field and are characterized
by reduced dimensions. A prototype of the SiPM + SciFi system was already tested on
DA$\Phi$NE (fibers were placed under the lower scintillator of the SIDDHARTA Kaon Monitor) \cite{ref:Bazzi,ref:Bazzi1}.
A second and more complex prototype, constituted of two layers of BCF-10 double cladded fibers,
free to rotate and read at both sides by Hamamatsu S10362-11-050-U SiPMs was recently tested. An excellent time resolution for kaons was achieved ($\sigma \sim 300$ ps) \cite{ref:nucl}.

%For what concerns the target system a half toroidal cryogenic target is under study, enclosed in a vacuum chamber. Kaons
%coming from $\phi$ decay will pass a degrader and then stop in the high density gaseous target, filled
%with $^3$He as a first step, $^4$He in a second phase. According to MC simulation about 20$\%$ of the negative kaons from DA$\Phi$NE should be stopped in the gas filling the target.

A half toroidal cryogenic target is under study, enclosed in a vacuum chamber, filled
with $^3$He as a first step, $^4$He in a second phase. According to MC simulation about 20$\%$ of the negative kaons from DA$\Phi$NE should be stopped in the gas filling the target. 
More outer layers of scintillating fibers opposite to the target cell would enable to clearly identify the K$^+$, and to perform a reconstruction of the inner trajectory of the
kaons. A similar target was recently installed in DA$\Phi$NE, for the SIDDHARTA \cite{ref:Curce1,ref:Curce2,ref:Curce3} experiment and our group will take advantage of the gained experience.

\section{Analysis of the KLOE data searching for $K^-$-$^4$He interactions}

Presently, we are performing dedicated Monte Carlo simulations, to study the performance of the AMADEUS setup. In parallel, we are analyzing the existing KLOE data (runs from 2002 to 2005). Indeed the KLOE drift chamber is mainly filled with $^4$He ($90\%$ helium $10\%$ isobutane) and the analysis of KLOE Monte Carlo showed, that about $0.1\%$ of kaons from DA$\Phi$NE should stop in the inner volume of the drift chamber. This represents a unique opportunity to check the reconstruction capability for $\Sigma$ and $\Lambda$ particles and for studying the hadronic interactions of $K^-$ in such an active target. Up to now data for a total luminosity of $1.8$ fb$^{-1}$ were analyzed \cite{ref:Cargn}. An excellent result was already achieved in reconstructing the $\Lambda$(1116) invariant mass, with a statistical error  $\sigma \sim 3$ keV; the systematics being presently under evaluation. The investigation of the $\Lambda$p and $\Lambda$d events, together with the study of the $\Lambda(1405)$ through the neutral decay channel $\Sigma^0\pi^0$ already gave excellent results, which are presently being finalized.

%\begin{figure}[htbp]
%\includegraphics[width=18.2pc, height=14pc]{3.pdf}
%\caption{Lateral view of the dedicated AMADEUS setup. The beam pipe is surrounded by scintillating fibers, read by silicon photomultipliers. The target %system is also represented.}
%\label{setup}
%\end{figure}

\acknowledgments
Part of this work was supported
by HadronPhysics I3 FP6 European Community
program, Contract No. RII3-CT-2004-506078;
the European Community-Research Infrastructure Integrating
Activity "Study of Strongly Interacting Matter"
(HadronPhysics 2, Grant Agreement No. 227431), and
HadronPhysics 3 (HP3), Contract No. 283286 under
the Seventh Framework Programme of EU.

\end{document}